\documentclass{aa} 
\usepackage{graphicx}
\usepackage{txfonts}
\usepackage{aalongtable}
\usepackage{longtable}

\usepackage{natbib}
\bibpunct{(}{)}{;}{a}{}{,}

\input epsf.sty

\begin{document}
\title{The existence of warm and optically thick dissipative coronae above accretion disks}
\author{A. R\'o\.za\'nska \inst{1}
        \and
        J. Malzac \inst{2,}\inst{3}
        \and
        R. Belmont \inst{2,}\inst{3}
        \and
        B. Czerny \inst{4}
        \and 
        P.-O. Petrucci \inst{5,}\inst{6}
        }
 \offprints{A. R\'o\.za\'nska}

\institute{Copernicus Astronomical Center, Bartycka 18, 00-716 Warsaw, Poland \\
          \email{agata@camk.edu.pl}
         \and
         Universit{\'e} de Toulouse, UPS-OMP, IRAP, Toulouse, France 
        \and 
         CNRS; IRAP; 9 Av. colonel Roche, BP44346, F-31028 Toulouse cedex 4, France
         \and 
         Center for Theoretical Physics, Al. Lotnik{\'o}w 32/46, 02-680 Warsaw, Poland
	\and 
	Universit{\'e} de Grenoble Alpes, IPAG, F-38000 Grenoble, France 
	\and  
	CNRS, IPAG, F-38000 Grenoble, France 
         }
\abstract
   {In the past years, several observations of AGN and X-ray binaries have 
  suggested the existence of a warm 
($T \sim 0.5-1$ keV) and optically thick ($\tau_{\rm cor} \sim 10-20$) corona covering the inner parts of the 
 accretion disk. These properties are directly derived from spectral fitting in UV to soft-X-rays using 
 Comptonization models. However, whether such a medium can be both in radiative and hydrostatic 
  equilibrium with an accretion disk is still uncertain.}
   {We investigate the properties of such warm,  optically thick coronae and put constraints on their
   existence.}
   {We solve the radiative transfer equation for grey atmosphere analytically in  a pure 
   scattering medium,   including local dissipation as an additional heating term in the warm corona. 
    The temperature profile of the warm corona is calculated 
   assuming it is cooled by Compton  scattering, 
   with the underlying dissipative disk providing photons to the corona.} 
   {Our analytic calculations show that  a dissipative thick ($\tau_{\rm cor}$ in the range 10-12)
   corona on the top of 
     a standard accretion disk 
   can  reach temperatures of the order of 0.5-1 keV  in its upper layers provided that the disk is passive.  
   But, in absence of strong magnetic fields, the requirement of a Compton cooled corona in hydrostatic 
   equilibrium in the vertical direction sets an upper limit on the Thomson optical depth  
   $\tau_{\rm cor}\lesssim 5$. We show  this value cannot be exceeded independently of the accretion
 disk parameters. However, magnetic pressure can extend this result to larger 
   optical depths.  Namely, a dissipative corona might have an optical depth 
   up to $\sim 20$ when  the  magnetic pressure is 100 
   times higher that the  gas pressure.}
   {The observation of warm coronae with Thomson depth larger than $\simeq 5$ puts tights constraints on the physics of the accretion disk/corona systems and requires either strong magnetic fields or vertical outflows to stabilize the system.}


\maketitle

\section{Introduction}
\label{sec:introduction}

Many successful models of the broad band spectra of accreting black holes contain
 a contribution from a moderately hot, optically thick layer on the top 
of the relatively colder accretion disk. Such a layer is frequently postulated when 
the broad band spectrum of an accreting system  resembles a power law 
in the soft X-ray band, with the photon index above~2. 

The standard theory of the \citet[][hereafter SS disk]{SS73} accretion disk model
predicts that  the  spectrum of an accretion disk is well modelled as a multi-color 
black body, showing an exponential cut-off at high 
frequencies connected with the maximum of the temperature in the 
disk atmosphere. Some exceptional quasars/active galaxies and specific spectral 
states of X-ray binaries selected for the determination of the black hole spin 
show such a thermal cut-off \citep{steiner2010}, but in  many cases the spectrum continues 
as a relatively steep/soft power law. 

The hard (2-100 keV) X-ray  spectra of radio quiet AGN is generally 
characterised by a flat power law shape sometimes cut-off around 100 keV 
\citep[][and references therein]{jourdain92,maisack93,perola02,ballantyne2014,malizia2014}
 and the presence of reflection components (iron line, reflection hump) 
is commonly observed \citep{pounds90}. The soft (below 2 keV) band is generally
 characterised by an excess with respect to the extrapolation of this
 hard X-ray power law. This is the so-called soft X-ray excess. 
 When fitted with a power law,  it shows a steep ($\Gamma > 2$) spectral shape. 
The origin of this component is still unknown. It could be equally
fitted by a blurred ionised reflection \citep{crummy2006}, 
a blurred ionised absorption \citep{gierlinski04}  or by thermal Comptonized 
emission in a moderately hot, optically thick layer possibly located on
the top of the colder accretion disk \citep{walter93,magdziarz98,done12,petrucci13}. 

The quasar composite spectra are well explained by such component, 
with the photon index $\Gamma \sim 2.5 $  \citep{laor1997,elvis12}. 
The whole class of  Narrow Line Seyfert 1 galaxies has also similar soft X-ray slopes. 
Such specific spectral element is observed in galactic sources in the 
Very High State and Intermediate State \citep[e.g.][]{gierlinski03}. 
The roughly power law shape of this component as well as the correlation observed between the UV and soft-X-ray bands suggests Comptonization 
as a mechanism responsible for this emission. Furthermore, observations show that 
the significant fraction of the bolometric luminosity of the accreting system 
($\sim 30 - 50 $\%) is carried out by this emitting layer
\citep{vasudevan14}. 

Usually, it is postulated that the hot medium responsible for this radiation forms a 
skin or a corona above the inner parts of the accretion disk. 
This skin or corona is optically thick in many models of specific objects, 
with Thomson optical depth of the order of 2 - 20.  
The observed slope of the soft X-ray spectrum does not determine the optical depth of 
the scattering medium since it depends on the Compton $y$ parameter, i.e. a combination of the 
optical depth and the temperature.
However, the fact that an unscattered disk component is not required in the fit or the direct detection of the 
turn-off imply rather low temperature and  high optical depth. 

Such solutions are usually discussed in the context of specific observational data. 
\citet{white1982} required an optically thick corona to explain the temporal and 
spectral properties of neutron star sources 4U 1822-37, 4U 2729+47 and 
Cyg X-3 \citep[see also][]{bayless10}, although recent papers \citep[e.g.][]{iaria13}
argue that the corona is very optically thin and the direct view to the neutron 
star in 4U 1822-37 is blocked by the outer rim of the accretion disk. 

\citet{magdziarz98} postulated the presence of a warm optically thick 
Comptonizing medium to fit the UV-Soft X-ray spectrum of NGC 5548. 
However,  they did not expect it to co-exist with the cold disk 
as a vertical layer; instead, they suggested this medium as a radial transition 
region between the cold outer disk and a hot inner flow.
\citet{zhang2000} modelled the spectrum of GRO J1655-40 with a warm layer at T=1.0 keV
with the optical thickness  $\tau=10$ located above a cold accretion disk.  
Indeed, they have considered three vertical layers: a cold disk, a warm skin and a hot corona.
\citet{janiuk01} required the presence of a warm corona of optical depth equal  to 
12 in order to fit the soft X-ray spectrum of a quasar/NLS1 object PG 1211+143.
\citet{zycki01} found coronal temperature $\sim 5$~keV and optical 
depth $\sim 3$ from their hybrid models of soft states of X-ray binaries:
GS 1124–68 and GS 2000+25, 
and higher values of optical depth, $\sim 10$, were implied
for some of the Very High State data sets. 
 \citet{petrucci13} for the Seyfert 1 galaxy Mrk 509,  found soft corona 
with temperature $\sim0.5 $~keV, and large optical depth $\sim 20$.

Monte Carlo models of hot coronae with large optical depth were
calculated by \citet{czerny03} in order to explain the broad-band spectra 
of quasar composite spectra, Ton S180, Mrk 359 and PG 1211+143, i.e. high 
accretion rate AGN. 
\citet{kubota04} obtained good spectral fits with the temperature of  the order 
of 10 keV and optical depth $\sim 2$ in the two data sets for VHS of XTE J1 550-564.  
\citet{jin12} successfully fitted the broad band spectra of 51 AGN with a model 
of accretion disk thermal emission,  a low temperature optically thick Comptonization and 
 a hot optically thin corona. In their fits the electron temperature in the thick 
corona was in the range  of 0.1 - 2 keV, and the optical depth from 4 to 40 in 
different objects. The model of the optically thick, low temperature corona surrounding
 the cold disk was also successfully applied to model the spectrum of a ULX source IC 342 
\citep{ebisawa03} although the model was considered
rather unphysical by the authors. 

Postulating an optically thick hotter medium sandwiching the colder disk inside 
is in apparent conflict with the results for the radiative transfer in the 
diffusion approximation.  A temperature inversion is usually only
 obtained in the optically thin zone, whereas the temperature rises towards the 
interior of the celestial body in the optically thick zone.
The best example is the solar corona. 
Thus the question arises of whether the cold disk would not heat up to reach the same temperature
as the optically thick part of the corona?
 
\begin{figure}[t!]
\vspace{1.5mm}
\begin{center}
\includegraphics[width=\columnwidth]{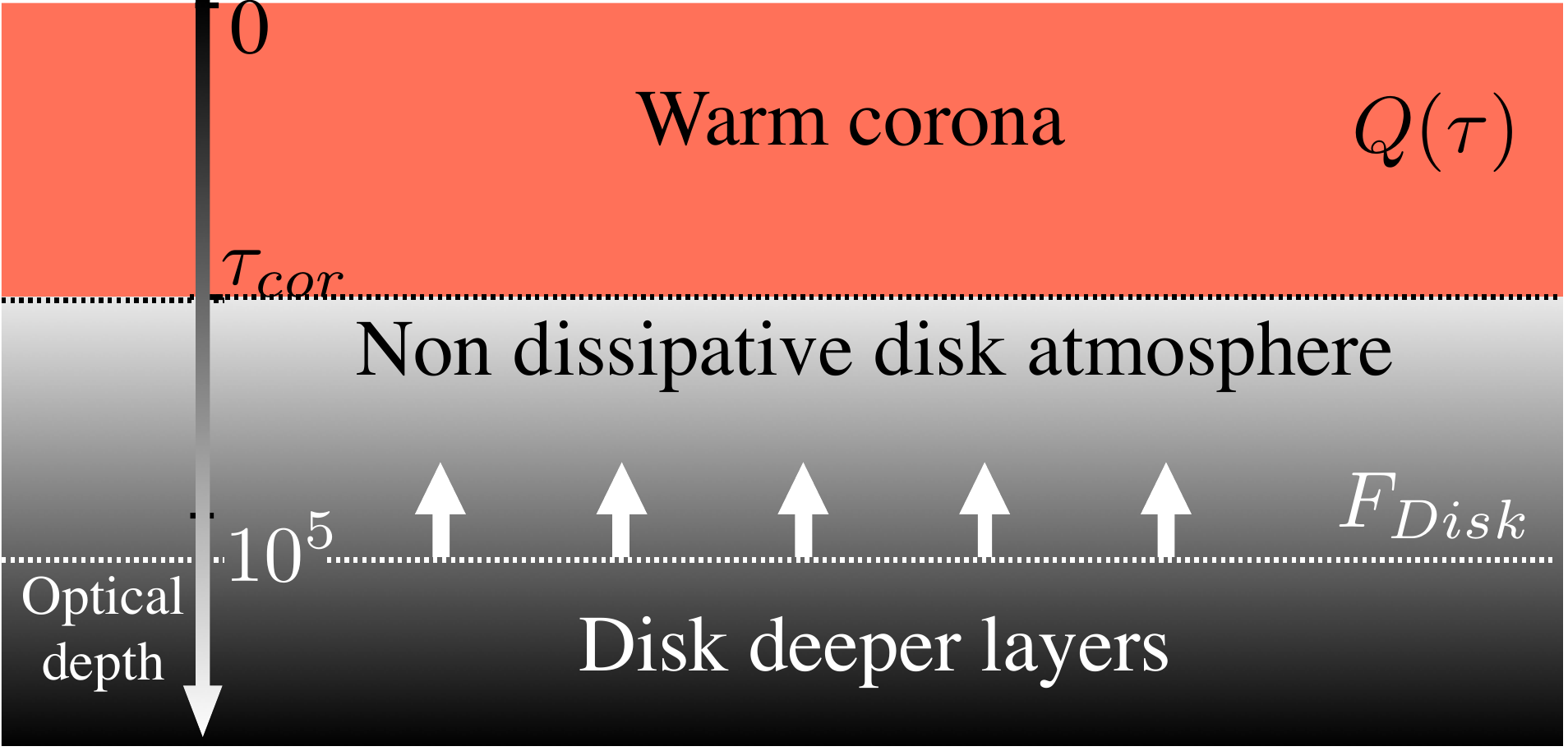}
\end{center}
\caption{Sketch of the dissipative slab corona atop the non-dissipative atmosphere of the accretion disk.}
\label{fig:coronadisk}
\end{figure}

In this paper, we address this question using a very simple analytical model. We consider the vertical structure of the accretion disk/corona that is sketched in Fig.~\ref{fig:coronadisk}, and investigate the radiative and pressure equilibrium of the upper layers of the accretion flow  as a function of the fraction of the accretion power that is dissipated in the corona.
We show, that the disk embedded in the hotter optically thick medium indeed does not 
heat up much more than in the case of an optically thin surrounding corona discussed 
in a basic paper of \citet{haardt93}.  
We check conditions for which an optically thick, Compton cooled zone can 
exist in hydrostatic equilibrium with a specified underlying colder disk. The paper is organized as follows:
in Sec.~\ref{sec:grey} we describe the solution of radiative transfer 
equation for the grey atmosphere with additional heating, in Sec.~\ref{sec:temp}
the temperature structure is presented, in Sec.~\ref{sec:hyd} the hydrostatic equilibrium 
is solved, and in Sec.~\ref{sec:comp} we show when corona is dominated
by Compton cooling. All results are discussed in Sec.~\ref{sec:dis}.

\section{Grey optically thick scattering medium with dissipation}
\label{sec:grey}

We are interested in the case of a grey atmosphere with pure scattering
(i.e. we neglect emission and absorption).
We assume that the atmosphere is  optically thick, and therefore adopt the Eddington 
approximation in the whole medium. We allow 
for an  additional dissipation which heats the corona.
For simplicity, we assume that the input energy rate per unit optical depth (and solid angle), 
 $Q$ is uniform from the surface ($\tau=0$) to the base of the corona ($\tau=\tau_{\rm cor}$). 
This corona is located above a cold accretion disk which itself dissipates energy
due to the accretion flow. The assumption that all dissipated 
energy is emitted in the form of radiation allows us to derive a simple fully 
analytical solution for the local vertical temperature  structure.  At the surface of the corona the escaping radiation flux $F_{\rm acc}^{\rm tot}$ is the sum of the flux  generated through internal dissipation inside the accretion disk, $F_{\rm disk}$, and that produced through dissipation in the corona, $F_{\rm cor}$:
\begin{equation}
 F_{\rm acc}^{\rm tot}=F_{\rm disk}+F_{\rm cor} \, .
 \end{equation}
The assumption that the heating of the corona is uniform gives:
\begin{equation}
F_{\rm cor}=4\pi Q\tau_{\rm cor} \, .
\label{eq:fc}
\end{equation}
Additionally, we define a convenient parameter:
\begin{equation} 
\chi = \frac{F_{\rm cor}}{F_{\rm acc}^{\rm tot}} \, ,
\label{eq:chi}
\end{equation} 
which is the fraction of the total accretion power that is dissipated in the corona.  

The frequency-integrated radiation transfer equation with an additional energy 
input $Q$ can be written as:
\begin{equation}
\mu {dI \over d\tau} = I - J -  Q \, ,
\label{eq:rt}
\end{equation}
where $\mu$ is the cosine of azimuthal angle. 
The optical depth, $\tau$ is measured downward, from the top of the corona 
toward the disk, $I(\mu,\tau)$ is the radiation intensity,  and  $J(\tau)$ is 
the mean intensity. 
The third term on the right hand side modifies the source function, $S$, and 
describes the increase in the photon energy due to the dissipation within 
the corona ($ S = J+ Q$). Following the standard Eddington approach, we can 
derive the solution of the radiative transfer equation by calculating its first moments, 
i.e. integrating over solid angles.  The zeroth moment gives:
\begin{equation}
H(\tau) = -\tau Q +  C_1 \, ,
\end{equation} 
The integration constant $C_1$ represents the Eddington flux at the top of the corona ($\tau=0$), which is, by definition,
$C_1 = F_{\rm acc}^{\rm tot}/4\pi$. Using Eqs.~\ref{eq:fc} and \ref{eq:chi}, the Eddington flux profile can then be written as:
\begin{equation}
H(\tau) = \frac{F_{\rm acc}^{\rm tot}}{4\pi}\left(1 - \frac{\chi \tau}{\tau_{\rm cor}}\right) \, .
\label{eq:eddf}
\end{equation} 
We note that at $\tau = \tau_{\rm cor}$ where the corona touches the cold disk, the downward flux corresponding to the illumination of the disk by the corona cancels out the upward flux of reprocessed/reflected radiation from the disk. Therefore the net radiation flux at $\tau_{\rm cor}$ is only that caused by internal dissipation in the cold disk $F_{\rm disk}$.

The first moment of the radiation transfer is:
\begin{equation}
K(\tau) =  \frac{F_{\rm acc}^{\rm tot}}{4\pi}\left(\tau - { \chi  \tau^2 \over 2 \tau_{\rm cor}} \right) + C_2 \,.
\label{eq:MSc}
\end{equation}
In the Eddington approximation, we accept $K = J/3$ at every optical depth
across the medium. 
In addition, at the corona surface we have only
outgoing radiation flux, as in standard stellar atmosphere, so we have the condition 
$J(0) = 2 H(0)$ which allows to
determine the constant $C_2$. Setting $\tau = 0$  in Eq.~\ref{eq:eddf} and~\ref{eq:MSc} we get
$C_2= F_{\rm acc}^{\rm tot}/6\pi$, so that 
the mean intensity as a function of the optical depth in the optically thick corona is 
 given by the expression:
\begin{equation}
J(\tau) =  \frac{3  F_{\rm acc}^{\rm tot}}{4\pi} \left(\frac{2}{3} +\tau - { \chi \tau^2 \over {2 \tau_{\rm cor} }}
           \right) \, . 
\label{eq:jot}
\end{equation}
Equations~\ref{eq:eddf}, \ref{eq:MSc}, and \ref{eq:jot} are valid only in the warm corona (i.e. for $\tau<\tau_{\rm cor}$) and they are largely independent of the underlying accretion disk structure. However the same formalism can be used to extend these solutions deeper in the disk atmosphere in order to investigate the effects of the presence of the corona on the upper layers of the accretion disk. 

\begin{figure}[t!]
\begin{center}
\includegraphics[width=\columnwidth]{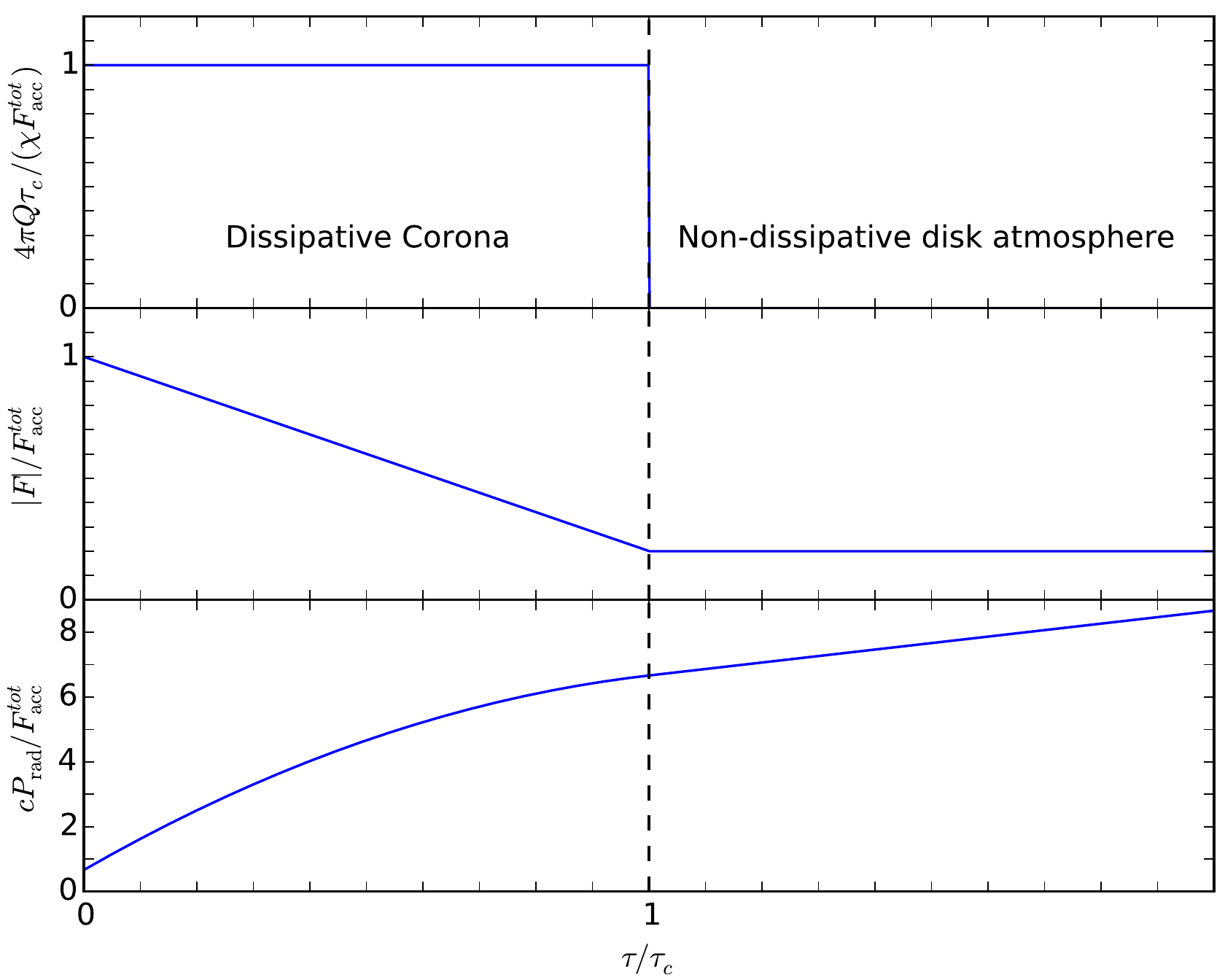}
\end{center}
\caption{Typical radiation structure of the accretion disk corona and upper disk atmosphere. The upper panel shows the assumed dissipation profile, the middle and lower  panel show the resulting radiation flux and pressure profiles, respectively.}
\label{fig:profilrad}
\end{figure}

As in the standard SS 1-D disk, all flux 
is generated close to the equatorial plane \citep{rozanska1999}, we can neglect dissipation in the disk atmosphere, and assume that all the disk flux is generated below this layer at deeper optical depth ($\tau \gtrsim 10^5$), see Fig.~\ref{fig:coronadisk}.  Here we consider only the properties of the upper non-dissipative atmosphere. For simplicity, we will not solve the  flux dissipation equation down to the midplane as this would not change our final result of temperature profile on the 
border of the warm corona and the disk.

The solution of Eq.~\ref{eq:rt} with $Q=0$, implies a constant Eddington flux below the corona. Then the  value of the flux at  $\tau_{\rm cor}$ sets the value of $H$ everywhere in the disk atmosphere:
\begin{equation}
H(\tau>\tau_{\rm cor})= \frac{F_{\rm disk}}{4\pi}= \frac{F_{\rm acc}^{\rm tot}}{4\pi}(1-\chi) \, .
\label{eq:hatm}
\end{equation}
The first moment of the radiation transfer then gives:
\begin{equation}
K(\tau>\tau_{\rm cor})=  \frac{F_{\rm acc}^{\rm tot}}{4\pi}(1-\chi)\tau+C_3 \, .
\end{equation}
The constant $C_3$ is determined from the condition of continuity of $K$ (and $J$) at $\tau_{\rm cor}$:  
\begin{equation}
C_3= \frac{F_{\rm acc}^{\rm tot}}{4\pi}\left(\frac{2}{3}+\frac{\chi}{2} \tau_{\rm cor}\right) \, .
\end{equation}
And finally the mean intensity in the disk atmosphere is:
\begin{equation}
J(\tau>\tau_{\rm cor})=  \frac{3F_{\rm acc}^{\rm tot}}{4\pi}\left[(1-\chi)\tau +\frac{\chi}{2}\tau_{\rm cor}+\frac{2}{3}\right] \, .
\label{eq:jda}
\end{equation}
We note that in absence of corona ($\chi=0$) this equation reduces to the standard mean intensity profile of grey atmospheres:
\begin{equation}
J_{\rm disk}= \frac{3F_{\rm disk}}{4\pi}\left(\tau+\frac{2}{3}\right) \, .
\end{equation}
In the presence of a warm corona, there is an additional component $J_{\rm cor}$ to the mean intensity of the disk that is due to the illumination of the atmosphere by the corona, $J=J_{\rm disk}+J_{\rm cor}$, which is:
\begin{equation}
J_{\rm cor}= \frac{3F_{\rm cor}}{4\pi}\left(\frac{\tau_{\rm cor}}{2}+\frac{2}{3}\right) \, .
\end{equation}
The typical vertical structure of the radiation properties of the corona/disk atmosphere system  is sketched in Fig.~\ref{fig:profilrad}.

 \begin{table}[b!]
\begin{center}
\caption{Parameters of the fiducial solutions illustrated in Figs.~\ref{fig:profil} and \ref{fig:profilrho}.  For all these models, the total accretion flux is set to  $F_{\rm acc}^{\rm tot} = 3.3 \times 10^{14}$ erg s$^{-1}$ cm$^{-2}$. The values of $\tau_{\rm cor}$ were choosen to be the maximum possible Thomson depth of a Compton cooled  corona  in hydrostatic equilibrium for the corresponding $\beta_{\rm m}$,  $\mathcal{G}$ and  $\chi$  (see Sect.~\ref{sec:comp}). $T_{\rm av}$ is the resulting average temperature of the corona estimated using Eq.~\ref{eq:ave}.}

\begin{tabular}{cccccc} 
\hline 
  Model  number   & $\beta_{\rm m}$ & $\mathcal{G}$ & $\chi$       & $\tau_{\rm cor}$ & $kT_{\rm av}$ (keV)  \\
\\      
 1  &                0                   &            0           &     0.98     &       5.21            &       3.91\\
 2  &              50			&             0            &    0.98     &       19.9            &       0.42  \\       
 3  &             50			&             5  	    &     0.98    &       8.76            &       1.68\\
 4  &             50			&             0	   &       0.4     &      15.9             &       0.23\\
 5  &            50			&             2 	   &	   0.4      &      7.08            &      0.88\\
 6  &              50			&             0           &      0.02   &         9.28         &       2.68 $\times 10^{-2}$\\
 \\
\hline
\label{tab:para}
\end{tabular}
\end{center}
\end{table}

\begin{figure}[t!]
\begin{center}
\includegraphics[width=\columnwidth]{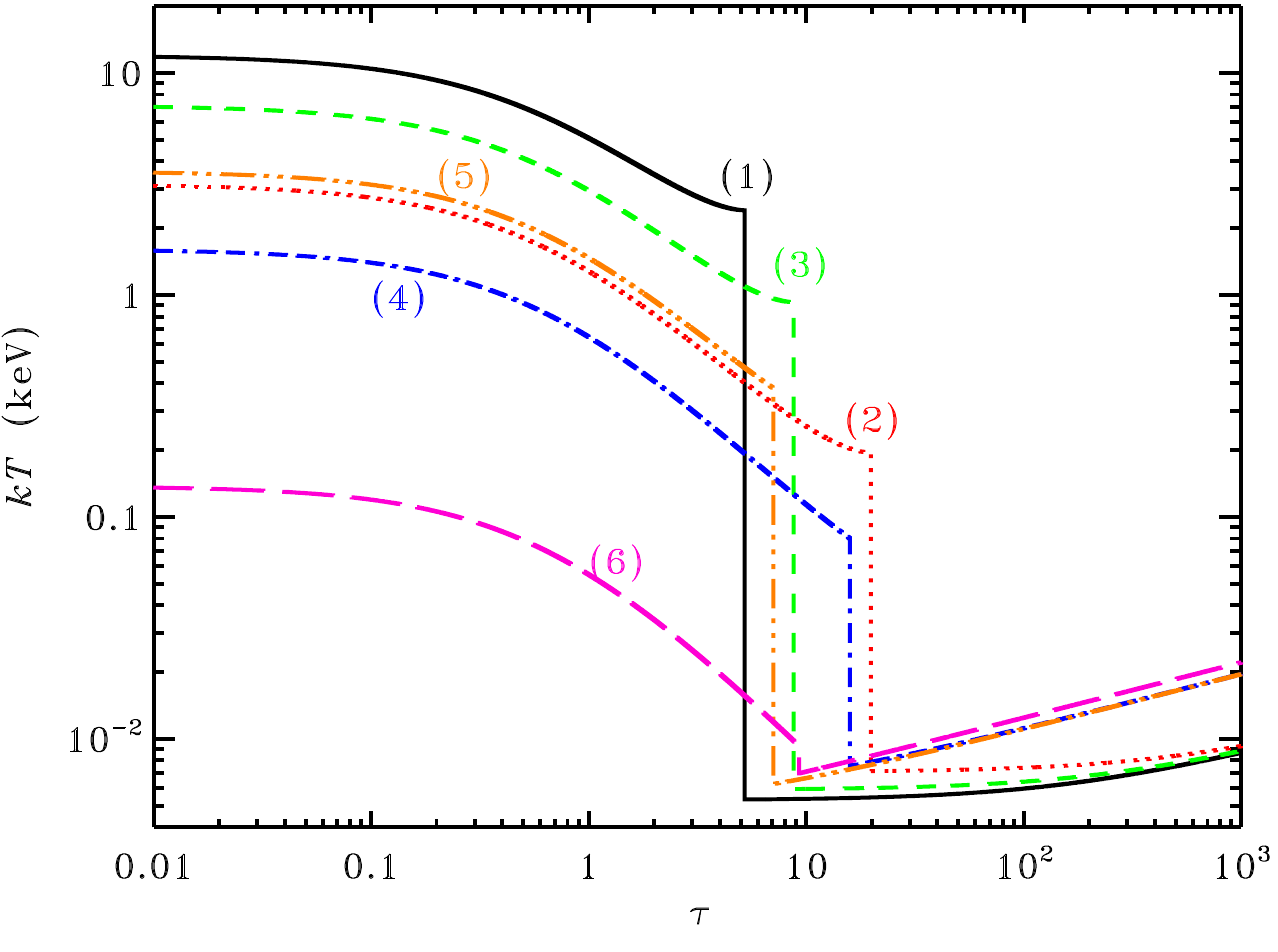}
\end{center}
\caption{Temperature profiles in the disk/corona system for the various values of the parameters $\chi$, $\tau_{\rm cor}$ of the six fiducial models reported in Table~\ref{tab:para}. Each curve is labelled with its model reference number, corresponding to that given in Table~\ref{tab:para}. }   \label{fig:profil}
\end{figure}

\section{The temperature profile}
\label{sec:temp}

The next step is to determine the temperature profile across the disk and corona 
using mean intensity field.  The temperature profile is derived assuming that matter is in equilibrium 
with the radiation field.  The resulting  temperature depends on the radiative cooling mechanism operating in the gas.

\subsection{Temperature profile in the warm corona}

The radiative transfer solution for grey atmosphere does not specify the temperature 
profile in a purely scattering medium which is the case of the soft corona, i.e.
for $\tau < \tau_{\rm cor}$.
However, we can obtain the temperature profile taking into account that the
scattered photons must cool the corona through Comptonization, in order
to have the thermal heating/cooling balance.
  
When cooling is dominated by inverse Compton scattering in the Thomson regime
 by a thermal population of sub-relativistic temperature, 
the balance between cooling and heating reads:
\begin{equation}
 J (\tau) {4 k T_{\rm cor}(\tau) \over m_{\rm e} c^2} = Q
\label{eq:bil}
\end{equation}
where $T_{\rm cor}$ is the corona temperature, $c$ -  the velocity of light, 
$k$ - the Boltzmann 
constant, and $m_{\rm e}$ - the electron rest mass. 

Since $J$ is an increasing function of the optical depth (see Eq.~\ref{eq:jot}) and $Q$ is assumed to be a constant,
the electron temperature decreases with 
$\tau$, showing a temperature inversion:
\begin{equation}
kT_{\rm cor}(\tau)={ \chi m_{\rm e} c^2 \over 12 \tau_{\rm cor}}  
       \left( {2 \over 3}+\tau - {\chi \tau^2 \over 2 \tau_{\rm cor}} 
         \right)^{-1}
\label{eq:tem}
\end{equation}

The temperature remains hot from the coronal surface down to $\tau=2/3$,
then it slowly decreases.  Depending on the amount of dissipated energy, the soft corona can be quite hot down to moderate optical depth. The vertical temperature profile of the soft corona above the cold disk is
determined by only  two parameters:  the total optical depth $\tau_{\rm cor}$
of the skin/corona, and the fraction $\chi$ of flux dissipated inside the corona to the 
total flux dissipated inside the disk and corona.
In~Fig.~\ref{fig:profil} we show examples of full temperature profiles 
for different values of these parameters.  Note, that this profile strictly comes from 
the radiation properties and does not imply that the hydrostatic equilibrium is 
satisfied in this multi-zone structure. We address this issue
in the section below.

\begin{figure}[h!]
\begin{center}
\hspace{-0.5cm}
\includegraphics[width=9cm]{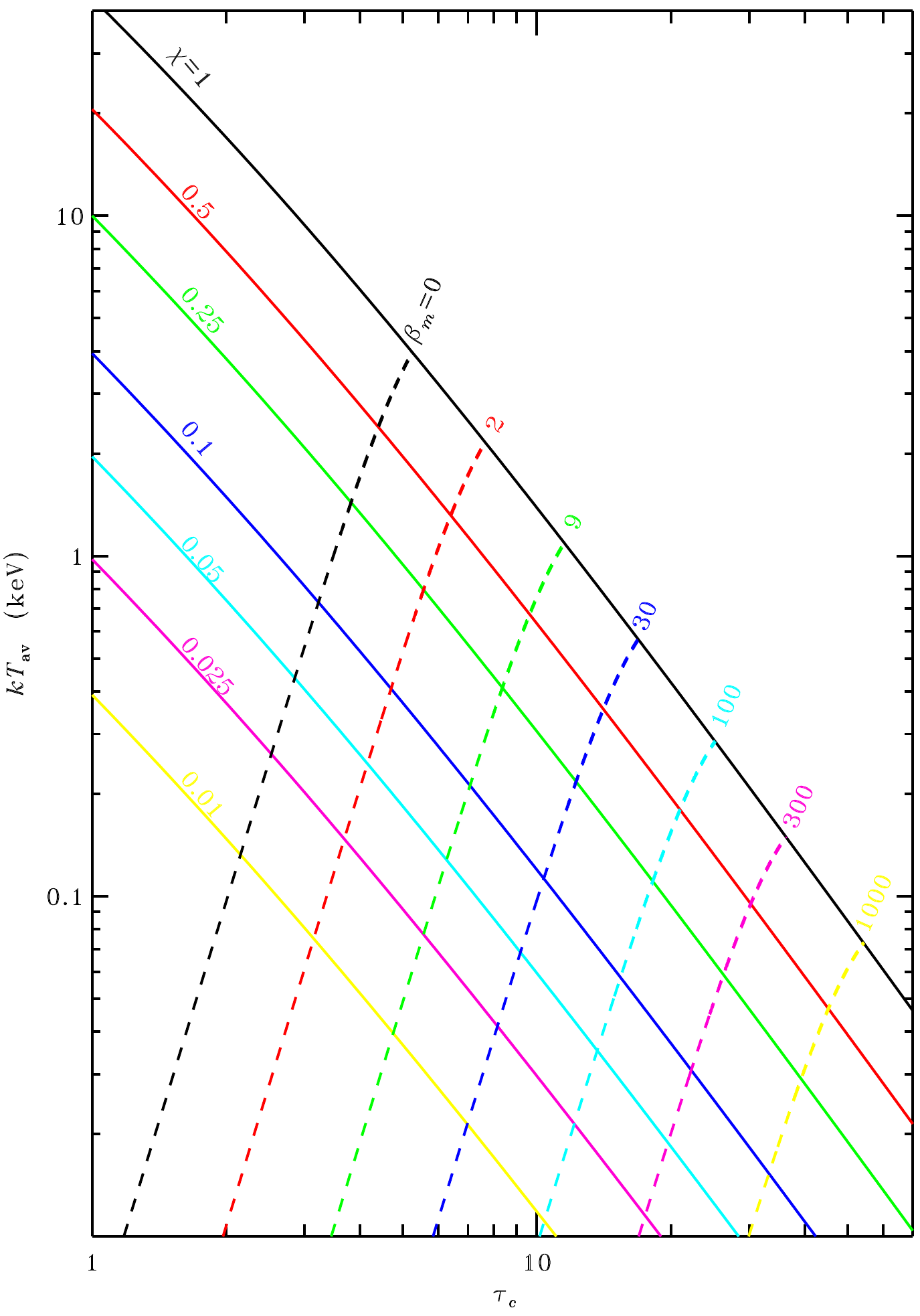}
\end{center}
\caption{Average coronal temperature vs. optical depth.  The full curves show the dependence of the average coronal temperature on the total optical depth of the corona for various values of the fraction of total power dissipated in the corona, $\chi$, as labelled. The dashed curves represent the  hydrostatic equilibrium solutions providing the largest possible Thomson depth of the Compton cooled corona  for a given  $\chi$ and magnetic to gas pressure ratio $\beta_{\rm m}$ (see Sect~\ref{sec:comp}). Each of the dashed curve shows the track of the solutions in the $kT_{\rm av}$-$\tau_{\rm cor}$ plane for a fixed value of  $\beta_{\rm m}$ (as labelled) when $\chi$ is varying.
}
\label{fig:tvstau}
\end{figure}

One-zone Comptonization models such as those used 
to derive the observational properties on the warm corona only 
constrain the average temperature of the skin,
without any constrains on the temperature stratification. 
Since, in our models temperature changes with optical depth, we should compare 
the value of the average temperature of the warm skin  weighted by its optical depth 
with that determined from observations:
\begin{equation}
T_{\rm av} = {1 \over \tau_{cor}}  \int_0^{\tau_{\rm cor}}  T_{\rm cor}(\tau) d \tau= {\chi m_e c^2 \over 12 k u \tau_{\rm cor}^2} \ln{\left(\frac{1+\frac{\chi}{u-1}}{1-\frac{\chi}{u+1}}\right)}\, ,
\label{eq:ave}
\end{equation}
where
\begin{equation}
u=\sqrt{1+\frac{4\chi}{3\tau_{\rm cor}}}.
\end{equation}
In the limit of large $\tau_{\rm cor}$, the average temperature given by Eq.~\ref{eq:ave} can be approximated within 10 percent (for $\tau_{\rm cor}>3$)  as:
\begin{equation}
kT_{\rm av} \simeq {\chi m_{\rm e}c^2 \over 12 \tau_{\rm cor}^2} \ln{\left(\frac{3\tau_{\rm cor}}{2-\chi}\right)} \, .
\label{eq:tav}
\end{equation}
The dependence of the average coronal temperature on $\tau_{\rm cor}$ is plotted in Fig.~\ref{fig:tvstau} for several values of the $\chi$ parameter. 
We can clearly see, that a warm skin, cooled by Comptonization, can be produced even for large values of the coronal optical depth provided  that most of the accretion power is dissipated in the warm corona. The coronal temperature however decreases with decreasing $\chi$ and increasing $\tau_\mathrm{cor}$.
 
As a  consequence, at small $\chi$, i.e. strong disk dissipation, a hot corona with pure 
scattering is only consistent with the most moderate observations (i.e. $\tau_{\rm cor}$ of the order of a few). 
From the energy equilibrium requirement
we can produce a layer of $T_{\rm cor} \sim 0.5-1$  keV  and large Thomson depth $\tau_{\rm cor }  \leqslant 15 $ only if most of the accretion power is dissipated in the layer. This appears to be consistent with the observational results of   \cite{petrucci13}  in the case of Mrk~509. These authors infer such parameters for the soft corona and also argue that the observed relative luminosity of the disk and soft corona implies that the disk is passive (i.e. $\chi\simeq 1$). The case of a passive disk provides the maximum achievable temperature for a given coronal depth. 
We note that some numerical  simulations of accretion disks also show a stronger dissipation in the outer layers of the disk \citep{ht2011}.

We also note that the method presented here is reasonably accurate. Our results can be compared for example to the Monte-Carlo simulations presented in  Malzac, Beloborodov \& Poutanen (2001), in the case of a slab corona  with Thomson depth of $\tau_{\rm cor}=3$ above a passive disk. They obtained an average temperature $\simeq 9$ keV  (see their figure~2)  which is in excellent agreement with the present results.  On the other hand for optically thin, coronae, the temperature becomes mildly relativistic, our approximations break down and the simple analytical model underestimates radiation cooling. For $\tau_{\rm cor}=0.5$ and $\chi=1$,  Eq.~\ref{eq:ave} gives $kT_{\rm av}\simeq103$ keV while  Malzac, Beloborodov \& Poutanen (2001) obtain  $kT_{\rm av}\simeq70$ keV with their detailed calculation.

\subsection{Temperature profile in the disk atmosphere}\label{sec:tempatm}

Deep in the  optically thick atmosphere of the disk ($\tau>\tau_{\rm cor}$) we can assume that radiation is fully thermalised ($J = \sigma T^4/\pi $, where $\sigma$ is the Stefan constant). Using Eq.~\ref{eq:jda},  we obtain the following temperature structure:
\begin{equation}
T_{\rm atm}^4= \frac{3F_{\rm acc}^{\rm tot}}{4\sigma} \left[(1-\chi)\tau+ \frac{2}{3}+\frac{\chi\tau_{\rm cor}}{2}\right]\, .
\label{eq:tther1}
\end{equation}
This expression can be rewritten as:
\begin{equation}
T_{\rm atm}^4=\frac{3}{4}T_{\rm disk}^4\left(\tau+\frac{2}{3}\right)+\frac{\pi J_{\rm cor}}{\sigma} \, .
\label{eq:tther2}
\end{equation}
The first term on the right hand side of Eq.~\ref{eq:tther2} corresponds to the standard temperature structure for fully thermalised grey atmosphere
in the Eddington approximation. $T_{\rm disk}$ is the disk effective temperature in absence of warm corona, calculated from the intrinsic disk flux $F_{\rm disk}$. The constant second term represents the increase in disk temperature due to the coronal illumination. 

Unlike $T_{\rm cor}$,  the temperature profile of the disk atmosphere, $T_{\rm atm}$, depends on the accretion flux $F_{\rm acc}^{\rm tot}$.
Using standard accretion disk theory, $F_{\rm acc}^{\rm tot}$ can be estimated as a function of the mass of the black hole, $M_{BH}$ ,and Eddington luminosity fraction, $\dot{m}=L/L_{\rm Edd}$, at a given radius $R=r GM_{\rm BH}/c^2$:
\begin{equation}
F_{\rm acc}^{\rm tot}\simeq 8\times 10^{26} \quad \frac{\dot m}{m} \frac{f}{r^{3}}   \quad {\rm erg \, s}^{-1} \, {\rm cm}^{-2} \, ,
\end{equation}
where $m=M_{\rm BH}/M_{\odot}$,  $f=2r_i(1-\sqrt{r_i/r})$ and $r_i$ is the inner radius of the disk expressed in gravitational radii. 

In Fig.~\ref{fig:profil}, the profile in the disk atmosphere was calculated for a black hole of mass
$M_{\rm BH} = 1.4 \times 10^8 M_{\odot}$ \citep{liu1983},
at 10 gravitational radii from the black hole, and 
at an accretion rate equal  to 2\% of the Eddington accretion rate, and we set $r_i=5$. For these parameters the accretion flux is $F_{\rm acc}^{\rm tot} = 3.3 \times 10^{14}$ erg s$^{-1}$ cm$^{-2}$.  As can be seen on Fig.~\ref{fig:profil}, the temperature profile in the upper layers of the disk flattens and departs from the standard grey atmosphere temperature profile due to the strong coronal illumination only when most of the power is dissipated in the warm corona.

 The absorption of coronal photons by the cold disk only slightly increases disk 
effective temperature which remains significantly lower than that of the corona. Due to additional dissipation in the soft corona, 
all temperature profiles show a strong temperature inversion in the disk/corona system. At the transition between the disk and corona there is a discontinuity due to the change in cooling mechanism. The amplitude of the temperature jump can be estimated as:
\begin{equation}
\left. \frac{T_{\rm atm}}{T_{\rm cor}} \right|_{\tau=\tau_{\rm cor}} \simeq 0.12 \left( \frac{\dot{m}}{m} \frac{f}{r^3}\right)^{1/4} \frac{\tau_{\rm cor}}{\chi}\left[\frac{2}{3}+\tau_{\rm cor}\left(1-\frac{\chi}{2}\right)\right]^{5/4} \, .
\end{equation}
This ratio remains lower than unity over a very broad range of black hole masses, mass accretion rates and disk radii. In practice, only when $\chi$ vanishes, the temperature of the disk at $\tau_{\rm cor}$ can become comparable or even hotter than that of  the corona. 

We note that in reality the change in cooling mechanism might not be as brutal as we have assumed here and the temperature discontinuity could  be smoothed. The computations of the disk/corona transitions are difficult
but it appears however that the temperature drop is always very sharp even if all the radiation processes are fully taken into account 
\citep{jurek2000,ballantyne2001,nayakshin2001,rozanska2002}.

\section{Hydrostatic equilibrium for the corona/disk system.}
\label{sec:hyd}

The question arises if  such a two-zone system can be in hydrostatic 
equilibrium. 
Here, we derive analytical formulae for the pressure profile in the disk-corona
system.
We use the standard equation of vertical hydrostatic equilibrium in geometrically thin disk \citep{madej2000}.
The total pressure, $P$, is the sum of the gas pressure $P_{\rm gas}$, the radiation pressure $P_{\rm rad}$,
and the magnetic pressure $P_{\rm mag}$. Locally, the total pressure 
has to balance the gravitational force:
\begin{equation}
{dP_{\rm gas} \over d{\tau}} + {dP_{\rm mag} \over d{\tau}}=  {1 \over \kappa_{\rm es}}  
{GM_{\rm BH} \over R^3}  z - {dP_{\rm rad} \over d{\tau}},
\label{eq:presseq}
\end{equation} 
where $G$ is  the gravitational constant, 
and $\kappa_{\rm es}$ -  the Thomson scattering cross section. 
In a general approach the total opacity should be taken into account, but for 
simplicity, we take only Thomson scattering into account and set $\kappa_{\rm es}=0.34$ cm$^{2}$~g$^{-1}$. In order to obtain analytical solutions, we will assume that  the warm corona is geometrically thin compared to the scale-height of the disk $Z_{\rm disk}$,  so that the vertical distance to the equatorial plane, $z$ , can be considered a constant $z=Z_{\rm disk}$.  This implies that the gravitational force is constant along the vertical direction inside the corona and the upper layers of the disk.

For the grey atmosphere, the radiation pressure gradient depends on the flux expressed 
in Eq.~\ref{eq:eddf} for the corona and Eq.~\ref{eq:hatm} for the disk atmosphere:
\begin{equation}
{dP_{\rm rad} \over d{\tau}} = { 4 \pi \over  c } H \,\,.
\label{eq:prad1}
\end{equation}

\subsection{Pressure and density profile in the warm corona}\label{sec:rhocor}

In the corona, the radiation pressure profile is obtained directly from Eq.~\ref{eq:jot}:
\begin{equation}
P_{\rm rad} =\frac{4\pi J}{3c}= { F_{\rm acc}^{\rm tot} \over c } \left(\tau + {2 \over 3}  - 
{\chi \tau^2 \over 2 \tau_{\rm cor}} \right) \,\,.
\label{eq:prad}
\end{equation}
We assume a uniform magnetic to gas pressure ratio $\beta_{\rm m}$:
\begin{equation}
P_{\rm mag}= {B_{\rm mag}^2 \over 8 \pi} = \beta_{\rm m} P_{\rm gas} \,\,.
\label{eq:mag}
\end{equation}
Solving the equation of hydrostatic equilibrium~(\ref{eq:presseq}), we find an expression for the 
gas pressure structure assuming, as a boundary condition, that $P_{\rm gas}(\tau=0) =0$: 
\begin{equation}
P_{\rm gas} =\frac{ F_{\rm acc}^{\rm tot}} {(1+\beta_{\rm m})c}  \left(
\mathcal{G}  \tau 
 + {\chi \tau^2 \over 2 \tau_{\rm cor}} \right) \, ,
\label{eq:pgas}
\end{equation} 
where the constant $\mathcal{G}$ represents the ratio of the pressure forces of the gas and magnetic field to that of radiation at the surface of the corona:
\begin{equation}
\left . \mathcal{G} = \left(\frac{dP_{\rm gas}}{d\tau}+\frac{dP_{\rm mag}}{d\tau}\right)/{\frac{dP_{\rm rad}}{d\tau}}\right|_{\tau=0}.
 \label{eq:gg2}
\end{equation}
The radiation pressure force dominates the support of the corona at all depths for $\mathcal{G}<1-2\chi$. $\mathcal{G}$ can also be expressed as:
\begin{equation}
\mathcal{G} = { {G M_{\rm BH}} \over R^3} {{c Z_{\rm disk}} \over {\kappa_{\rm es}  F_{\rm acc}^{\rm tot}}}-1\, .
\label{eq:gg}
\end{equation}
 The first term in the right hand side of Eq.~\ref{eq:gg} is the ratio of the gravitational to radiation pressure force at the surface of the corona. For a corona in hydrostatic equilibrium this ratio is necessarily larger than unity and consequently $\mathcal{G} \ge 0$. 
The half of the disk thickness $Z_{\rm disk}$ is a crucial parameter  
which controls the hydrostatic equilibrium in the vertical 
direction \citep{roza1999}.  The case  $\mathcal{G}=0$ gives the minimum disk thickness, for which the 
total disk pressure can be balanced by the gravitational force:
\begin{equation}
Z^{\rm min}_{\rm disk}= {\kappa_{\rm es}\, F_{\rm acc}^{\rm tot} R^3 \over G 
M_{\rm BH} \, c} = \frac{3}{2} \frac{GM_{\rm BH}}{c^2}f \dot{m}\,.
\end{equation} 
For a thinner disk, the matter will be outflowing from the system \citep{witt97}. 

The density profile clearly depends on the assumed disk geometrical thickness $Z_{\rm disk}$,
and we can derive this density from the equation of state as:
\begin{equation}
\rho = {{\mu m_{\rm H}} \over kT_{\rm cor}} {F_{\rm acc}^{\rm tot} \over (1+\beta_{\rm m}) c}  \left(
      \mathcal{G} \tau + {{\chi \tau^2} \over 2 \tau_{\rm cor}} \right)\,\,,
\label{eq:rho}
\end{equation}
where $\mu$ is a mean molecular weight assumed to be 0.5, 
and $m_{\rm H}$ is the mass of hydrogen atom. The density increases with $\tau$ and from Eq.~\ref{eq:rho} we see that setting the disk parameter $\mathcal{G}=0$ (or equivalently $Z_{\rm disk}=Z^{\rm min}_{\rm disk}$) minimises the density in the corona.  This is illustrated in Fig.~\ref{fig:profilrho} which displays examples of  density profiles obtained for the parameters listed in Table~\ref{tab:para}.

\subsection{Pressure and density profiles in the disk atmosphere}\label{sec:diskatm}
The pressure and density profiles in the disk atmosphere below the corona ($\tau>\tau_{\rm cor}$) can be estimated in a similar way.
The radiation pressure profile is given by Eq.~\ref{eq:jda}:
\begin{equation}
P_{\rm rad}=  \frac{F_{\rm acc}^{\rm tot}}{c}\left[(1-\chi)\tau +\frac{\chi}{2}\tau_{\rm cor}+\frac{2}{3}\right].
\label{eq:pda}
\end{equation}
The pressure equilibrium equation ~(\ref{eq:presseq}) is solved using Eq.~\ref{eq:hatm} and assuming continuity of pressure at the disk/corona transition: 
\begin{equation}
P_{\rm gas}=  \frac{F_{\rm acc}^{\rm tot}}{(1+\beta_{\rm m})c}\left[(\mathcal{G}+\chi)\tau-\frac{\chi}{2}\tau_{\rm cor}\right] \, .
\end{equation}
The density profile follows:
\begin{equation}
\rho = {{\mu m_{\rm H}} \over kT_{\rm atm}} {F_{\rm acc}^{\rm tot} \over (1+\beta_{\rm m}) c} \left[(\mathcal{G}+\chi)\tau-\frac{\chi}{2}\tau_{\rm cor}\right]
\,\,,
\label{eq:rhoatm}
\end{equation}

\begin{figure}[t!]
\begin{center}
\hspace{-0.5cm}
\vspace{-1.0cm}
\includegraphics[width=9cm]{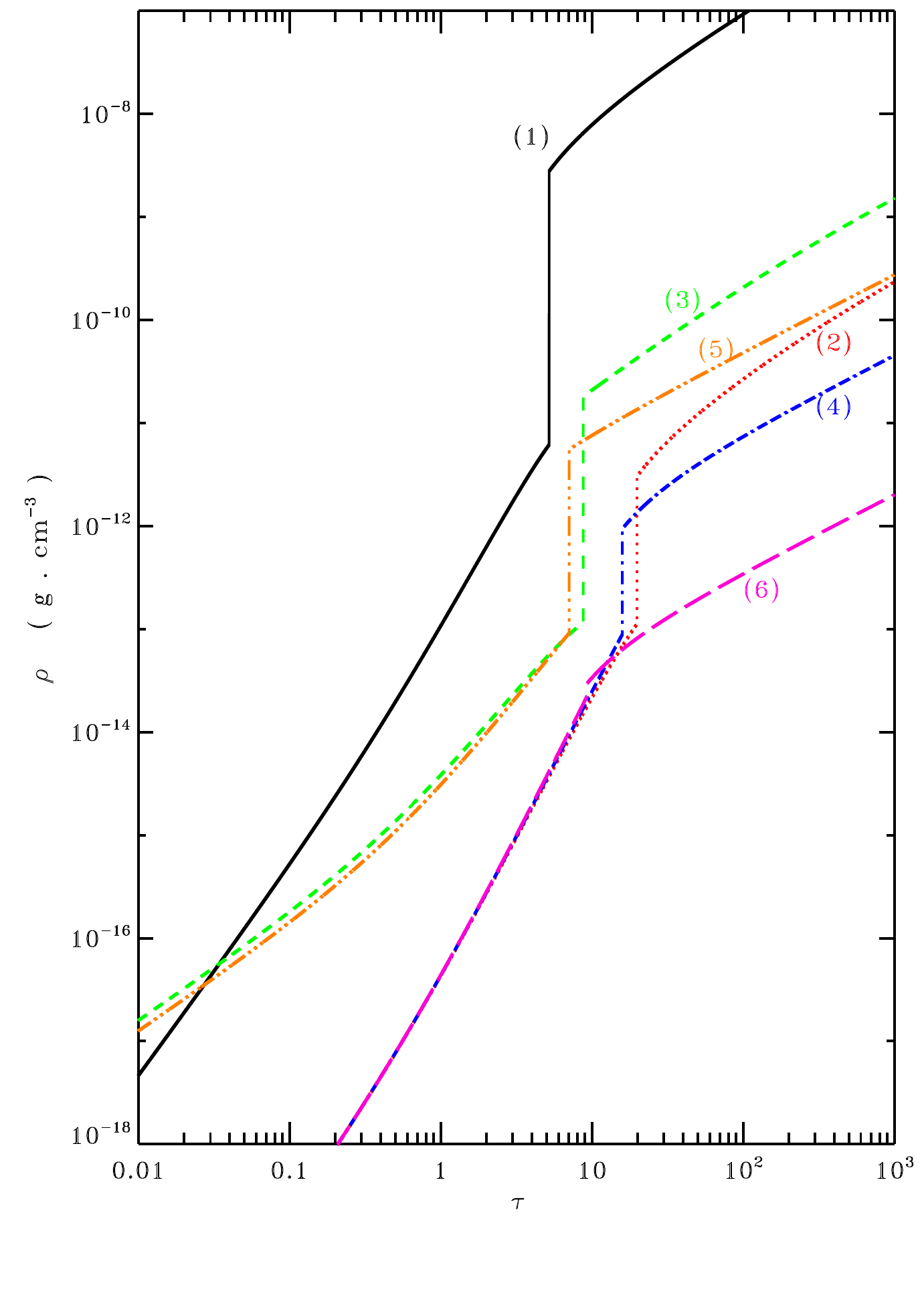}
\end{center}
\caption{Density profile  around the disk corona transition for the 6 fiducial models detailed in Table~\ref{tab:para}.  The curves are labelled by their model number.}
\label{fig:profilrho}
\end{figure}

Fig.~\ref{fig:profilrho} shows some examples of density profiles around the disk transitions for our fiducial value of the total accretion flux. In all cases,  we observe a discontinuity in density at $\tau_{\rm cor}$ which has the same amplitude as the temperature jump  discussed in the  Sect.~\ref{sec:tempatm}. Due to the temperature jump and the condition of pressure equilibrium at the disk/corona transition, the disk atmosphere tends to be much denser than the corona.

\subsection{Limitations}

In Sect.~\ref{sec:hyd} we have estimated the properties of a warm corona and disk atmosphere in pressure equilibrium.  Our proposed  treatment presents the advantage of being very simple. The drawbacks of this simplicity are some limitations that we now briefly discuss.  

First, althought the effects of the disk on the corona and disk atmosphere is taken into account via the $\mathcal{G}$ parameter, the present approach does not allow us to guarantee that there is indeed a disk solution below the atmosphere that both has the required $\mathcal{G}$ parameter and connects smoothly to the atmosphere. A full calculation of the vertical stratification of the disk down to the mid-plane, including also dissipation, would be required in order to obtain such self-consistent solutions and calculate $Z_{disk}$ from first principles.  

Also we note that for the density profiles with the lowest densities in the corona, our assumption of constant gravity may be inaccurate. Indeed for these profiles the scale height of the corona:
\begin{equation}
H_{\rm cor}\sim\int_{\tau_{\rm cor}/2}^{\tau_{\rm cor}}(\kappa_{\rm es}\rho)^{-1}d\tau \, , 
\end{equation}
can be be comparable, or even larger than $Z_{\rm disk}$. In this case  the gravity at the surface of the corona is significantly larger than at the bottom. 
 Taking into account the height dependent gravity would require a numerical resolution of the equilibrium which is out of the scope of this paper, but  we can anticipate its effects. Indeed, if we assume that the disk/corona transition is located at a height $Z_{\rm disk}$, the increased gravity force in the upper corona will necessitate a larger pressure in order to sustain the equilibrium, and as a consequence the coronal density will also be increased compared to our current estimates. In particular this effect may affect our results for small $Z_{\rm disk}$ (or small $\mathcal{G}$), which may underestimate the pressure and and density in the corona by a factor of up to a few.  

Finally our calculation of the pressure equilibrium assumes that the opacity is dominated by electron scattering. If absorption becomes important, both pressure and density will be reduced compared to our simple estimates. We have checked a posteriori that for the fiducial models presented in Figs.~\ref{fig:profil} and~\ref{fig:profilrho} the Kramer free-free absorption opacity $\kappa_{\rm ff}\simeq 6 \times 10^{22} \rho T^{-7/2}$ cm$^{2}$ g$^{-1}$  is negligible  compared to $\kappa_{\rm es}$ both in the the corona an the atmosphere of the disk. For these models our estimated  pressure profiles are not affected by the approximation of a pure scattering medium. We stress however that since even in the disk atmosphere the medium is only weakly absorbing,  the assumption of fully thermalised radiation used to infer the temperature profile of the atmosphere may break down close to the disk/corona transition, making the transition much more gradual than our simplified calculation suggest.  

A detailed investigation of all these issues is deferred to future works.

\section{Constraints from the requirement of a Compton cooled corona}
\label{sec:comp}

In the previous sections we have determined the temperature, pressure and density profile of the corona under the assumption that the dominant cooling mechanism is Compton scattering. This assumption was motivated by observational results based on the modelling the coronal  emission with Comptonization models. We now have to determine the parameter regimes for which this assumption remains valid.
Besides Compton cooling, the most efficient cooling mechanism is expected to be bremsstrahlung which must remain negligible compared to Compton 
cooling. Here we estimate the ratio of Compton  cooling rate 
$\Lambda_{\rm C} = 16 \pi kT / (m_{\rm e} c^2) \, \rho \, \kappa_{\rm es} J(\tau) $ 
(in erg/s/cm$^3$) to the
bremsstrahlung  cooling rate $\Lambda_{\rm B} = B \, \rho^2 T^{1/2}$, 
 where $B = 6.6 \times 10^{20}$  CGS units. This ratio must remain larger than unity
across the  soft corona.  Using  Eqs.~\ref{eq:jot}  and \ref{eq:tem} we get the condition:
\begin{equation}
{\Lambda_{\rm C} \over \Lambda_{\rm B}}  = A \frac{1+\beta_{\rm m}}{\left(\tau_{\rm cor}/\chi\right)^{3/2}}
\left( {2 \over 3} + \tau - { \chi \tau^2 \over 2 \tau_{\rm cor} }\right)^{-\frac{1}{2}}\left(\mathcal{G}\tau+\frac{\chi\tau^2}{2\tau_{\rm cor}}\right)^{-1}\ge 1,
\label{eq:cond}
\end{equation}
with the constant
$A = \sqrt{k m_{\rm e}} \, c^2 \kappa_{\rm es} / (\sqrt{12} B \, \mu \, m_{\rm H})\simeq 57$.
The above ratio is a decreasing function of optical depth, therefore
the condition~\ref{eq:cond} is verified in the whole corona if it verified  
at $\tau=\tau_{\rm cor}$. The maximum possible Thomson depth for a Compton cooled corona is obtained by setting the condition that the depth $\tau_{\rm cor}$  corresponds to the transition between Compton dominated to bremsstrahlung dominated regions, i.e. by solving $\left. \Lambda_{\rm C} / \Lambda_{\rm B} \right|_{\tau=\tau_{\rm cor}}=1$. The coronal optical depth of the fiducial models of Table~\ref{tab:para} leading to the profiles presented in Fig.~\ref{fig:profil} and \ref{fig:profilrho} were determined in this way and correspond to the deepest possible Compton corona for the given set of $\chi$, $\beta_{\rm m}$ and $\mathcal{G}$.  

From Eq.~\ref{eq:cond}, we also see that  the case of $\mathcal{G} = 0$, or equivalently $Z_{\rm disk}=Z^{\rm min}_{\rm disk}$, gives the most favorable condition to have a purely Compton cooled corona because it corresponds to the minimum of the gas pressure and density. 
Therefore, in order to find regimes where our assumptions fail, 
it is enough to calculate this ratio at the base of corona for $\mathcal{G}=0$: 
\begin{equation}
\left. {\Lambda_{\rm C} \over \Lambda_{\rm B}} \right|_{\tau=\tau_{\rm cor}}
 \cong \frac{\sqrt{8}A (1+\beta_{\rm m}) }{\tau_{\rm cor}^3 \sqrt{2/\chi-1}} \ge 1.
\label{eq:base}
\end{equation}
We stress that this equation represents the necessary condition for the Compton cooling dominance and 
does not depend on any disk parameters.
 
\begin{figure}
\begin{flushright}
\hspace{-5.7mm}
\vspace{-3.0mm}
\includegraphics[width=9.5cm]{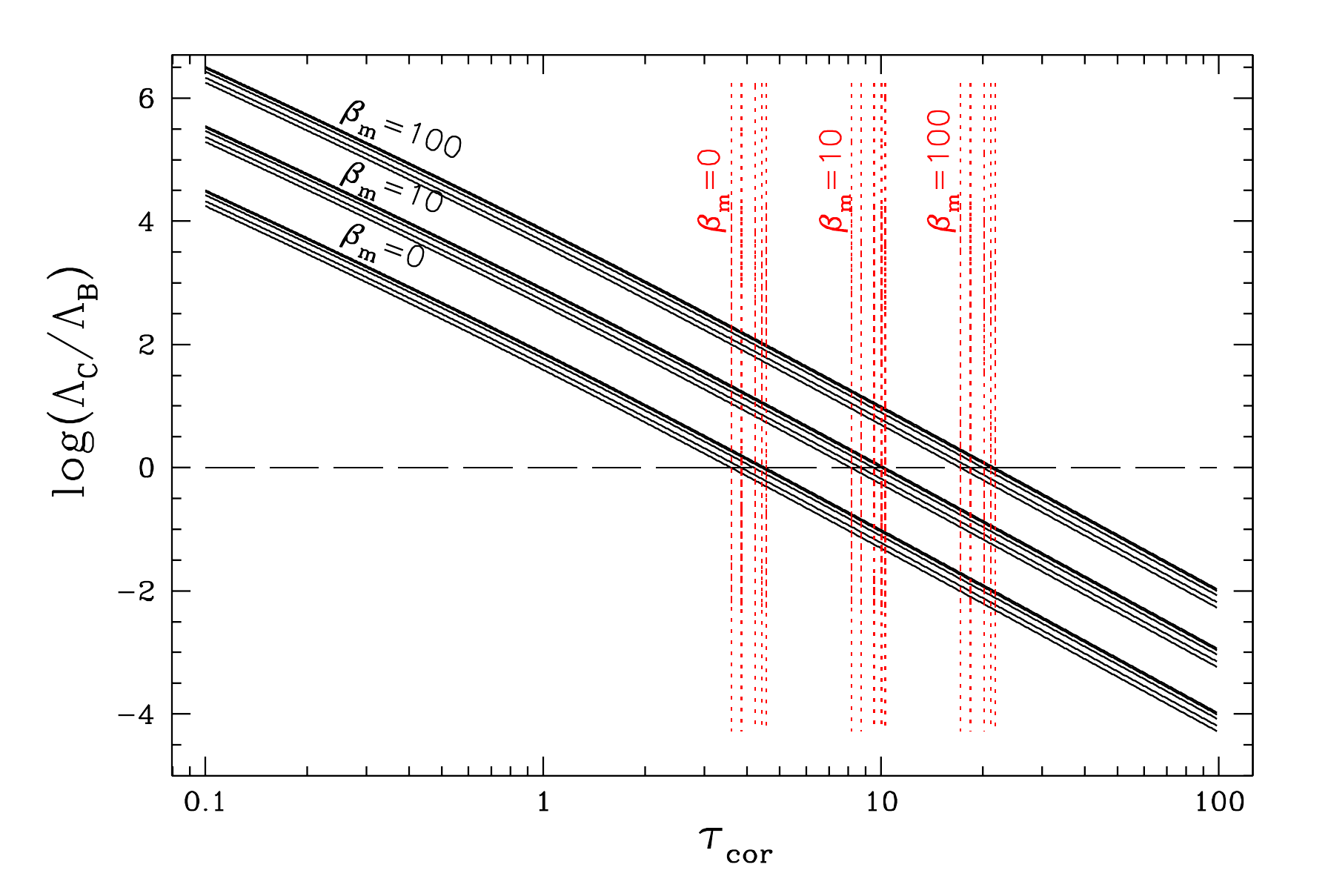}
\end{flushright}
\caption{The importance of Compton scattering over the bremsstrahlung 
at the base of corona as a function of $\tau_{\rm cor}$ (Eq.~\ref{eq:base}).
Solid lines in each package are computed for various values
of $\chi=0.18, 0.26, 0.40, 0.57, 0.67$ and 0.86. Each package is calculated for 
three different values of magnetic pressure: $\beta_{\rm m}=0, 10$ and 100. 
The horizontal dashed line represents the case where $ \Lambda_{\rm C} / \Lambda_{\rm B} =1 $,
while vertical dotted lines mark values of coronal optical depth for 
which it happens. 
 }
\label{fig:compt}
\end{figure}

Fig.~\ref{fig:compt} shows the Compton to bremsstrahlung cooling ratios as a function 
of coronal optical depth for three different  values of magnetic pressure in the case $\mathcal{G}=0$. For each value 
of $\beta_{\rm m}$ we consider various values of $\chi$ to show the 
marginal influence of this parameter. 
We see that the maximum possible Thomson depth of the corona in the case $\mathcal{G}=0$ provides an absolute upper limit on the depth of the corona. This upper limit can be estimated as:
  \begin{equation}
 \tau_{\rm cor}\simeq 5.4 \quad (1+\beta_{\rm m})^{1/3}(2/\chi-1)^{-1/6} \, .
 \label{eq:taumax}
 \end{equation}
 The corresponding average temperature of this `deepest' possible corona can then be estimated using Eq.~\ref{eq:tav}:
\begin{equation}
kT_{\rm av}\simeq  4  \,  \frac{\chi^{\frac{2}{3}}(2-\chi)^{\frac{1}{3}}}{(1+\beta_{m})^{\frac{2}{3}}} \left[1-1.2\times 10^{-4}  \ln{\frac{\chi^{-1}(2-\chi)^7  }{(1+\beta_{\rm m})^{2}}}\right] \, {\rm keV.} 
 \end{equation}

Eq.~\ref{eq:taumax} shows that when magnetic pressure is zero, bremsstrahlung becomes the dominant emission process
as soon as $\tau_{\rm cor} \gtrsim 5$. Such solutions are therefore inconsistent with 
an unmagnetised, static, Compton dominated, hot, and optically thick corona. Relaxing 
one of these constraints might produce consistent solution. For instance, the corona might not be in static equilibrium. The case of outflowing 
coronae \citep{witt97} in the frame of our model will be investigated in a future work.

Alternatively, additional magnetic pressure helps the gas pressure to
balance gravity. Hence it produces solutions with lower density, i.e. 
with lower bremsstrahlung cooling rate. 
 Fig.~\ref{fig:tvstau} maps the iso-contours of $\chi$ and $\beta_m$ for the deepest possible corona, in the $\tau_{\rm cor}$ vs $kT_{\rm av}$ plane.   
This diagram allows one to estimate the minimum value of the magnetic field pressure ratio required in order to produce a corona with a given optical depth and temperature.   
For instance, the  warm corona recently observed in Mrk 509 by Petrucci et al. (2013), with 
optical thickness  $\tau_{\rm cor} \sim 15$ and temperature $T_{\rm av}\simeq0.5$ keV, implies magnetic to gas pressure ratio
$\beta_{\rm m}>30$.  For $M_{\rm BH}=1.4 \times 10^8 M_\odot$, a distance of 10 gravitational radii, and an accretion rate equals to 
2\% of the Eddington accretion rate, this implies that the average magnetic field in the corona must be larger than $10^3$ G.

Since the calculations are based on the assumption of $\mathcal{G} = 0$, 
they provide the necessary conditions for the Compton cooling condition to be satisfied.
If Compton cooling dominance is not obtained, no change of $Z_{\rm disk}$ will save the situation, 
and if Compton cooling dominance is obtained, there is always a range of 
$Z_{\rm disk} \geqslant Z ^{\rm min}_{\rm disk}$, for which the hydrostatic equilibrium is sustained, and 
density is low enough for Compton cooling to dominate over bremsstrahlung cooling.
We note however that in the case of minimal density in the corona ($\mathcal{G}=0$) our approximation of constant gravity force in the vertical direction can break down (see Sec.~\ref{sec:rhocor}). As a consequence our simple calculations may underestimate the density and bremsstrahlung cooling rate. Relaxing the assumption of constant gravity may reduce the maximum possible Thomson depth of the corona and/or require even stronger magnetic fields to maintain the dominance of Compton cooling.   
 
\section{Discussion and Conclusion}
\label{sec:dis}

 In this paper, we put constraints on the existence of 
a warm, dissipating, optically thick, and Compton cooled corona 
in hydrostatic equilibrium with a cold accretion disk.  We neglect synchrotron photons, thermal conduction and 
ionisation in the warm skin, but we have checked that those processes 
are not important comparing to Comptonization. In our computations of the hydrostatic 
equilibrium in the vertical direction, the radiation pressure component is fully taken into account,  contrary 
to numerical simulations by \citet{schnit2013,uzdensky13}.

Our simple analytical solution for the warm and dissipative corona 
above the cold disk shows that a stable temperature inversion is possible 
in the optically thick case, contrary to the intuitive
expectations based on the diffusion approximation. This is due to the 
fact that the solution of purely scattering atmosphere does not 
specify the temperature. It only gives the 
radiation density, which rises with the optical depth, as expected. 
The temperature in the warm corona is determined 
a posteriori, from Compton cooling balance, and the temperature increases
toward the warm skin surface.
We have shown that such corona can  reach  temperatures of  0.5-1 keV 
for assumed values of constant dissipation in the skin of moderate
optical depth ($\tau_{\rm cor}<10$). The most extreme parameters e.g. a coronal temperature 0.5 keV at optical depth 15-20, which 
is observed in the case of Mrk~509 \citep{petrucci13}, can also reproduced provided that the disk is passive (i.e. almost all of the accretion power is dissipated in the corona).

Nevertheless, if this zone is in hydrostatic equilibrium with the cold accretion
disk, the maximum optical depth of the corona cannot exceed $\sim 5$ without any additional 
magnetic pressure.This upper limit is independent of the disk parameter. 
Higher optical depth of the warm 
skin is possible if the gas pressure is lowered, by magnetic pressure, or 
possibly by mass outflow. 
In this paper, we illustrate only the first case i.e. non zero value 
of magnetic field strength. When, the ratio of 
magnetic pressure to the gas pressure is 100, the  maximal optical depth of the 
warm corona is around 20, which is consistent with  some observations.

We conclude, that in the absence of magnetic pressure, 
additional dissipation in the outer layer is able to 
heat up the corona, but the requirement of hydrostatic balance with the disk, puts
strong limit on the coronal optical thickness. 
This limit is independent on the accretion disk parameters, i.e. its accretion rate 
and the mass of the central black hole.  

In this context, the X-ray observations of optically thick ($\tau_{\rm cor}$$>$ 5), warm coronae have a strong implication for the disk/corona system:
either strong magnetic fields or vertical outflows are 
required to stabilise the system. What is more, the simple conditions discussed 
in this paper are the minimum requirements for the existence of the thick corona, and further
modelling of the disk/corona interaction may likely impose even more stringent constraints for
the existence of such medium.

\begin{acknowledgements}
This research was conducted within the scope of the HECOLS International Associated
Laboratory, supported in part by the Polish NCN grant
DEC-2013/08/M/ST9/00664. AR and BC were supported by NCN grants No. 2011/03/B/ST9/03281, 2013/10/M/ST9/00729, and by Ministry of Science and Higher Education grant W30/7.PR/2013.  
They have received funding from the European Union Seventh Framework Program (FP7/2007-2013) under 
grant agreement No.312789. This research has also received fundings from PNHE in France, and from the french Research National Agency: CHAOS project ANR-12-BS05- 0009 (http://www.chaos-project.fr).  JM and POP also acknowledge fundings from CNRS/PICS.
\end{acknowledgements}

\bibliographystyle{aa}
\bibliography{refs}

\end{document}